\documentclass[preprint]{sig-alternate} 

\usepackage{ifthen}
\usepackage{listings}
\usepackage{amsmath}
\usepackage{color}
\usepackage{graphicx} 
\usepackage{colortbl}
\usepackage{float}
\usepackage{xspace}
\usepackage{multirow}
\usepackage{syntax}
\usepackage{url}
\usepackage{tikz}
\usetikzlibrary{arrows,positioning,shapes,shadows}

\makeatletter
\def\@copyrightspace{\relax}
\makeatother

\let\othelstnumber=\thelstnumber
\def\createlinenumber#1#2{
    \edef\thelstnumber{%
        \unexpanded{%
            \ifnum#1=\value{lstnumber}\relax
              #2%
            \else}%
        \expandafter\unexpanded\expandafter{\thelstnumber\othelstnumber\fi}%
    }
    \ifx\othelstnumber=\relax\else
      \let\othelstnumber\relax
    \fi
}

\newcommand{\internalcomment}[1]{
\begin{small}
\noindent
\textbf{** Comment:} {\color{red} #1} \\
\noindent
\textbf{** End comment}
\end{small}
}
\renewcommand{\internalcomment}[1]{}

\newcounter{todos}

\newcommand{\annotation}[2]{%
  \textsf{\textcolor{#2}{$^{[\thetodos]}$}}%
  \marginpar{%
    \framebox[0.9\marginparwidth][t]{%
      \parbox[t]{0.85\marginparwidth}{%
        \raggedright\scriptsize{
          \textcolor{#2}{\thetodos: #1}%
        }}}}%
  \stepcounter{todos}%
}

\newcommand{\TODO}[1]{\annotation{#1}{red}}
\newcommand{\NOTE}[1]{\annotation{#1}{blue}}

\newcommand{\figpath}{.}

\newcommand{\level}{level\xspace}
\newcommand{\levels}{levels\xspace}

\newcommand{\hcir}{HC IR\xspace}
\newcommand{\llvmir}{LLVM IR\xspace}

\newcommand{\ciao}{Ciao\xspace}
\newcommand{\ciaopp}{CiaoPP\xspace}
\newcommand{\tool}{tool\xspace}

\newcommand{\ial}{Ciao assertion language\xspace}

\newcommand{\isa}{ISA\xspace}

\newcommand{\lsem}{\mbox{$\lbrack\hspace{-0.3ex}\lbrack$}}
\newcommand{\rsem}{\mbox{$\rbrack\hspace{-0.3ex}\rbrack$}}
\newcommand{\sem}[1]{\lsem #1 \rsem}
\newcommand{\Inten}{\mbox{$ I$}}
\newcommand{\p}{{\sem{P}}}

\newtheorem{auxexample}{Example}[section]

\newenvironment{Example}{\begin{auxexample}\em }{\ $\Box$\end{auxexample}}

\definecolor{lightgrey}{rgb}{0.95,0.95,0.95}

\renewcommand{\NOTE}[1]{}
\renewcommand{\TODO}[1]{}

\newcommand{\secbeg}{}
\newcommand{\secend}{}

\newcommand{\kbd}[1]{\mbox{\tt #1}}
\newcommand{\nt}[1]{\mbox{\it #1}}

\def\imp{\hbox{${\tt \ :\!-\ }$}}

\lstdefinelanguage{xc}{
  keywords={int,if,return, struct, void},
  keywordstyle=\bf
}[keywords]

\lstdefinelanguage{ciao}{
  keywords={regtype, trust, num, list, rsize, var, resource},
  keywordstyle=\bf
}[keywords]

\lstdefinelanguage{llvm}{
  keywords={getelementptr, label, br, i1, i32},
  basicstyle=\ttfamily,
  keywordstyle=\bf,
  xleftmargin=.02\textwidth,
  frame=l,
  numbers=left,
  numberstyle=\scriptsize
}[keywords]

\lstdefinelanguage{hcir}{
  keywords={trust},
  basicstyle=\ttfamily,
  keywordstyle=\bf,
  frame=l,
  xleftmargin=.02\textwidth, 
  numbers=left,
  numberstyle=\scriptsize
}[keywords]

\definecolor{darkgreen}{rgb}{0,0.75,0}
\definecolor{darkblue}{rgb}{0,0,0.75}
\definecolor{lightgreen}{rgb}{0.85,1,0.85}
\definecolor{lightblue}{rgb}{0.85,0.85,1}
\definecolor{grey85}{rgb}{0.45,0.45,0.45}
\definecolor{lightblue2}{rgb}{0.75,0.75,1}

\definecolor{lightblue3}{rgb}{0.70,0.70,0.92}
\definecolor{magenta}{rgb}{1,0,1}
\definecolor{sienna}{rgb}{1,0,1}
\definecolor{maroon4}{rgb}{0.55,0.11,0.38}
\definecolor{chocolate}{rgb}{0.82,0.41,0.12}
\definecolor{red4}{rgb}{0.55,0,0}
\definecolor{darkred}{rgb}{0.75,0,0}

\newcommand{\upsymb}[1]{{\Large \textsuperscript{#1}}}
\renewcommand{\email}[1]{{\normalsize \texttt{#1}}}
\begin{document}

\title{Towards Energy Consumption Verification via Static Analysis}

\numberofauthors{8} 
\author{
\alignauthor
P.~Lopez-Garcia
       \titlenote{Spanish Council for Scientific Research
         (CSIC).}\upsymb{\dag}\\
       \email{pedro.lopez@imdea.org}
\alignauthor
R.~Haemmerl\'{e}
       \titlenote{IMDEA Software Institute, Madrid, Spain.}\\
       \email{remy.haemmerle@imdea.org}
\alignauthor 
M.~Klemen \upsymb{\dag} \\
       \email{maximiliano.klemen@imdea.org}
\and 
\alignauthor 
U.~Liqat \upsymb{\dag}\\
       \email{umer.liqat@imdea.org}
\alignauthor 
M.~Hermenegildo 
\upsymb{\dag} \titlenote{Universidad Polit\'{e}cnica de Madrid (UPM).}\\
       \email{manuel.hermenegildo@imdea.org}
}
\additionalauthors{Additional authors: John Smith (The Th{\o}rv{\"a}ld Group,
email: {\texttt{jsmith@affiliation.org}}) and Julius P.~Kumquat
(The Kumquat Consortium, email: {\texttt{jpkumquat@consortium.net}}).}
\date{30 July 1999}

\maketitle

\begin{abstract}
  In this paper we leverage an existing general framework for resource
  usage verification and specialize it for \emph{verifying} energy
  consumption specifications of embedded programs.  Such
  specifications can include both lower and upper bounds on energy
  usage, and they can express intervals within which energy usage is
  to be certified to be within such bounds. The bounds of the
  intervals can be given in general as functions on input data
  sizes. Our verification system can prove whether such energy usage
  specifications are met or not. It can also infer the particular
  conditions under which the specifications hold. To this end, these
  conditions are also expressed as intervals of functions of input
  data sizes, such that a given specification can be proved for some
  intervals but disproved for others.  The specifications themselves
  can also include preconditions expressing intervals for input data
  sizes.  We report on a prototype implementation of our approach
  within the \ciaopp~system for the XC language and XS1-L
  architecture, and illustrate with an example how embedded software
  developers can use this tool,
  and in particular for determining values for program parameters that
  ensure meeting a given energy budget while minimizing the loss in
  quality of service.


\end{abstract}

\secbeg
\vspace*{3mm}
\section{Introduction}
\label{introduction}
\secend

In an increasing number of applications, particularly those running on
devices with limited resources, it is very important and sometimes
essential to ensure conformance with respect to specifications
expressing non-functional global properties such as energy
consumption, maximum execution time, memory usage, or user-defined
resources. For example, in a real-time application, a program
completing an action later than required is as erroneous as a program
not computing the correct answer. The same applies to an embedded
application in a battery-operated device (e.g., a portable or
implantable medical device, an autonomous space vehicle, or even a
mobile phone) if the application makes the device run out of batteries
earlier than required, making the whole system useless in practice.

In general, high performance embedded systems must control, react to,
and survive in a given environment, and this in turn establishes
constraints about the system's performance parameters including energy
consumption and reaction times. Therefore, a mechanism is necessary in
these systems in order to prove correctness with respect to
specifications about such non-functional global properties.

To address this problem we leverage an existing general framework for
resource usage analysis and 
verification~\cite{resource-verif-iclp2010-short,resource-verif-2012-short}, and
specialize it for \emph{verifying} energy consumption specifications of
embedded programs. As a case study, we focus on the energy
verification of embedded programs written in the XC
language~\cite{Watt2009} and running on the XMOS XS1-L architecture
(XC is a high-level C-based programming language that includes
extensions for communication, input/output operations, real-time
behavior, and concurrency).  However, the approach presented here can
also be applied to the analysis of other programming languages and
architectures. We will illustrate with an example how embedded
software developers can use this tool, 
and in particular for determining values for program parameters that
ensure meeting a given energy budget while minimizing the loss in
quality of service.

\section{Overview of the Energy Verification Tool}
\label{sec:overview}

\begin{figure*}
  \centering


\pgfdeclarelayer{background}
\pgfdeclarelayer{foreground}
\pgfsetlayers{background,main,foreground}

\tikzstyle{source}=[draw, draw=cyan!80!black!100, fill=cyan!20, text width=8em, font=\sffamily,
    thick,
    minimum height=2.5em,drop shadow]
\tikzstyle{tool}=[draw=green!50!black!100, fill=green!40, text width=5em, font=\sffamily, 
    thick,
    text centered, 
    chamfered rectangle, chamfered rectangle angle=30, chamfered rectangle xsep=2cm]
\tikzstyle{midresult}=[draw, fill=white!40, text width=7.6em, font=\sffamily,
    thick,
    rounded rectangle,
    minimum height=1em,drop shadow]
\tikzstyle{warnresult}=[color=orange!50!black!100, align=center, fill=orange!40, text width=5em, font=\sffamily, 
    thick,
    rounded corners=2pt,
    minimum height=1em,drop shadow]
\tikzstyle{errresult}=[color=red!80!black!100,  align=center, fill=red!20, text width=5em, font=\sffamily, 
    thick,
    rounded corners=2pt,
    minimum height=1em,drop shadow]
\tikzstyle{inforesult}=[align=center, fill=black!10, text width=5em, font=\sffamily, 
    thick,
    rounded corners=2pt,
    minimum height=1em,drop shadow]
\tikzstyle{okresult}=[color=green!50!black!100,  align=center, fill=green!40, text width=5em, font=\sffamily, 
    thick,
    rounded corners=2pt,
    minimum height=1em,drop shadow]
\tikzstyle{certresult}=[color=blue!50!black!100, fill=blue!40, text width=5em, font=\sffamily, 
    thick,
    rounded corners=2pt,
    minimum height=1em,drop shadow]
\tikzstyle{coderesult}=[color=blue!50!black!100, fill=blue!40, text width=5em, font=\sffamily, 
    thick,
    rounded corners=2pt,
    minimum height=1em,drop shadow]

\begin{tikzpicture}[>=latex,scale=1]
  \node (assertions) [source] {
    \textbf{Assertions}\\[2ex] \footnotesize
    \# pragma check\\
    \# pragma trust\\
    ...
  };
  \path (assertions)+(0,-6em) node (codeXC) [source] {
    \textbf{XC Code}\\[2ex]\footnotesize
     int f(int arg)\{\\
    ...\\[2ex]~
    
  };

  \path (assertions)+(0,7.5em) node (model) [source, align=center, text width=8.7em, minimum height=1.1cm ] {
    \textbf{Energy Model}\\
    };
  
\path (assertions)+(10.5em,4em) node (translator) [tool] {HC IR Translator};
\path (translator)+(0,-8em) node (compiler) [tool] {XC Compiler };
\path (translator)+(9.5em,0) node (statana) [tool] {Static Analysis};
\path[color=black] (statana)+(10em,0em) node (true) [midresult] {\#pragma true};
\path (statana)+(0em,-8em) node (comparator) [tool] {Static Comparator};
\path[color=blue] (comparator)+(10em,-2.5em) node (check) [midresult] {\#pragma check};
\path[color=red] (comparator)+(10em,0em) node (false) [midresult] {\#pragma false};
\path[color=green!50!black] (comparator)+(10em,2.5em) node (checked) [midresult] {\#pragma checked};

\path (true)+(9em,0em) node (inferred) [inforesult] {Inferred};
\path (false)+(9em,0em) node (cterror) [errresult] {Disproved};
\path (check)+(9em,0em) node (verifwarn) [warnresult] {Unproved};
\path (checked)+(9em,0) node (verified) [okresult] {Proved};

\path [draw, thick, ->] (model.east) -- node [] {} (translator.north west) ;
\path [draw, thick, ->] (assertions.east) -- node [] {} (translator.south west) ;
\path [draw, thick, ->] (codeXC) -- node [] {} (compiler.west) ;
\path [draw, thick, ->] (compiler) -- node [] {} (translator) ;
\path [draw, thick, ->] (translator.south east) -- node [] {} (comparator.north west) ;
\path [draw, thick, ->] (translator.east) -- node [] {} (statana.west) ;
\path [draw, thick, ->] (comparator.east) -- node [] {} (check.west) ;
\path [draw, thick, ->] (comparator.east) -- node [] {} (false.west) ;
\path [draw, thick, ->] (comparator.east) -- node [] {} (checked.west) ;
\path [draw, thick, ->] (statana) -- node [] {} (true) ;
\path [draw, thick, ->] (statana) -- 
(comparator) ;
\path [draw, thick, ->] (true) -- node [] {} (inferred) ;
\path [draw, thick, ->] (check) -- node [] {} (verifwarn) ;
\path [draw, thick, ->] (false) -- node [] {} (cterror) ;
\path [draw, thick, ->] (checked) -- node [] {} (verified) ;

\path [color=black] (statana.north)+(0,1.5em) node (preprocessor) {\sffamily \normalsize Energy Consumption Analysis \& Verification Tool};

\path [color=black] (assertions.north)+(0,1.5em) node (program) {\sffamily  \bf Program};

\begin{pgfonlayer}{background}
  \path (translator.north west)+(-0.5,1.0) node (g) {};
  \path (check.south east)+(0.5,-0.5) node (h) {};
  
  \path[fill=yellow!20,rounded corners, draw=black!50, densely dashed] (g) rectangle (h);
\end{pgfonlayer}

\begin{pgfonlayer}{background}
  \path (assertions.north west)+(-0.1,0.8) node (g) {};
  \path (codeXC.south east)+(0.1,-0.1) node (h) {};
  
  \path[source] (g) rectangle (h);
\end{pgfonlayer}

\end{tikzpicture}


\caption{Energy consumption verification tool using \ciaopp.}
\label{fig:analysis-verif-hcir}.
\end{figure*}
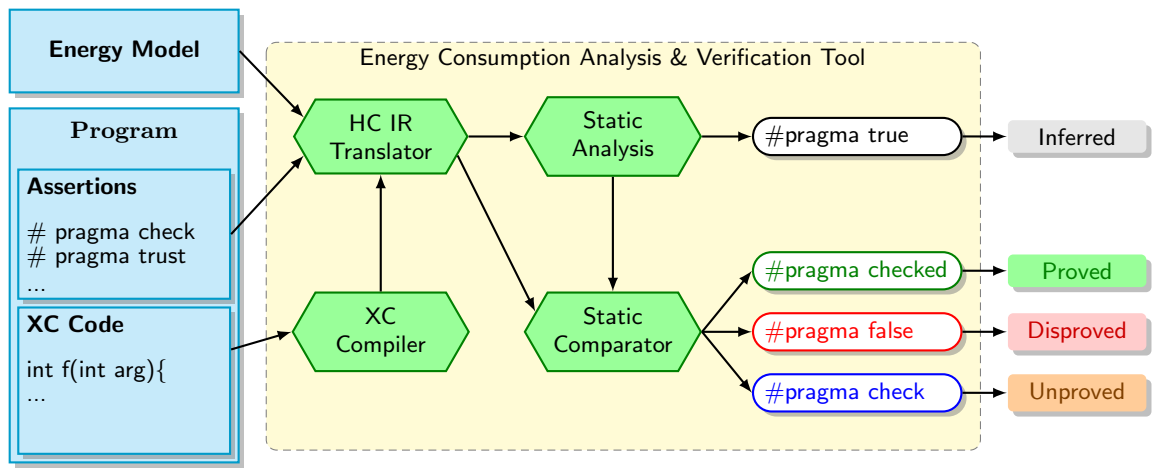

In this section we give an overview of the prototype tool for
\emph{energy consumption verification} of XC programs running on the
XMOS XS1-L architecture, which we have implemented within the
\ciaopp system~\cite{ciaopp-sas03-journal-scp-short}. As in previous
work~\cite{isa-energy-lopstr13-final-short,decomp-oo-prolog-lopstr07-short}, we
differentiate between the \emph{input language}, which can be XC
source, \llvmir~\cite{LattnerLLVM2004}, or Instruction Set
Architecture (ISA) code, and the \emph{intermediate semantic program
  representation} that the \ciaopp core components (e.g., the
analyzer) take as input. The latter is a series of connected code
blocks, represented by Horn Clauses, that we will refer to as
``\hcir'' from now on.  We perform a transformation from each
\emph{input language} into the \hcir and pass it to the corresponding
\ciaopp component.  The main reason for choosing Horn Clauses as the
intermediate representation is that it offers a good number of
features that make it very convenient for the
analysis~\cite{decomp-oo-prolog-lopstr07-short}. For instance, it supports
naturally Static Single Assignment (SSA) and recursive forms, as will
be explained later.
In fact, there is a current trend favoring the use of Horn Clause
programs as intermediate representations in analysis and verification
tools~\cite{hcvs14}.

Figure~\ref{fig:analysis-verif-hcir} shows an overview diagram of the
architecture of the prototype tool we have developed.  Hexagons
represent different tool components and arrows indicate the
communication paths among them.

The \tool takes as input an XC source program (left part of
Figure~\ref{fig:analysis-verif-hcir}) that can optionally contain
assertions in a C-style syntax. As we will see later, such assertions
are translated into \ciao assertions, the internal 
representation used in the \ciao/\ciaopp system.

The energy specifications that the tool will try to prove or disprove
are expressed by means of assertions with \texttt{check} \emph{status}.
These specifications can include both lower and upper bounds on energy
usage, and they can express intervals within which energy usage is to
be certified to be within such bounds. The bounds of the intervals
can be given in general as functions on input data
sizes. Our tool can prove whether such energy usage specifications are
met or not. It can also infer the particular conditions under which
the specifications hold. To this end, these conditions are also
expressed as intervals of functions of input data sizes, such that a
given specification can be proved for some intervals but disproved for
others.

In addition, assertions can also express trusted information
such as the energy usage of procedures that are not developed yet, or
useful hints and information to the \tool.  In general, assertions
with status \texttt{trust} can be used to provide information about the
program and its constituent parts (e.g., individual instructions or
whole procedures or functions) to be trusted by the analysis system,
i.e., they provide base information assumed to be true by the
inference mechanism of the analysis in order to propagate it
throughout the program and obtain information for the rest of its
constituent parts.

In our tool the user can choose between performing the analysis at the
\isa or \llvmir \levels (or both). 
We refer the reader to~\cite{entra-d3.2.4-isa-vs-llvm-short} for an
experimental study that sheds light on the trade-offs implied by
performing the analysis at each of these two \levels, which can help
the user to choose the level that fits the problem best.
\ \\

The associated \isa and/or \llvmir
representations of the XC program are generated using the xcc
compiler. Such representations include useful metadata.
The \emph{\hcir translator} 
component 
(described in Section~\ref{sec:llvm-ciao-translation}) produces the internal
representation used by the tool, \hcir, 
which includes
the program and possibly specifications and/or trusted information
(expressed in the
\ial~\cite{assert-lang-disciplbook-short,hermenegildo11:ciao-design-tplp-short}).

The tool performs several tasks:

\begin{enumerate}

\item Transforming the \isa and/or \llvmir into 
  \hcir. Such transformation
  preserves the resource consumption semantics, in the sense that the
  resource usage information inferred by the tool is applicable to the
  original XC program.

\item Transforming specifications (and trusted information) written as
  C-like assertions into the \ial.  

\item Transforming the energy model at the \isa
  \level~\cite{Kerrison13},  
expressed in JSON format, into the
  \ial. Such assertions express the energy consumed by individual
  \isa instruction representations, information which is required by the analyzer in order to 
   propagate it during the static analysis of a program 
  through code segments, conditionals, loops, recursions,
   etc., in order to infer analysis information (energy consumption
   functions) for higher-level entities such as procedures, functions, or
   loops in the program.

\item In the case that the analysis is performed at the \llvmir level,
  the \emph{\hcir translator} component produces a set of \ciao
  assertions expressing the energy consumption corresponding to
  \llvmir block representations in \hcir. Such information is produced
  from a mapping of \llvmir instructions with sequences of \isa
  instructions and the \isa-\level energy model. The mapping
  information is produced by the \emph{mapping tool} that was first
  outlined in~\cite{entra-d3.2} (Section 2 and Attachments D3.2.4 and
  D3.2.5) and is described in detail in~\cite{Georgiou14}.

\end{enumerate}

Then, following the approach described
in~\cite{isa-energy-lopstr13-final-short}, the \ciaopp parametric static
resource usage analyzer~\cite{resource-iclp07-short,NMHLFM08,plai-resources-iclp14-short}
takes the \hcir, together with the assertions which express the energy
consumed by \llvmir blocks and/or individual \isa instructions, and
possibly some additional (trusted) information, and processes them,
producing the analysis results, which are expressed also using \ciao
assertions. 
Such results include energy usage functions (which depend on input
data sizes) for each block in the \hcir (i.e., for the whole program
and for all the procedures and functions in it.).  The analysis can
infer different types of energy functions (e.g., polynomial,
exponential, or logarithmic). The procedural interpretation of the
\hcir programs, coupled with the resource-related information
contained in the
(\ciao) assertions, together allow the resource analysis to infer static
bounds on the energy consumption of the \hcir programs that are
applicable to the original \llvmir and, hence, to their corresponding
XC programs.  
Analysis results are given using the assertion language, to ensure
interoperability and make them understandable by the programmer.

The verification of energy specifications 
is performed by a specialized
component which compares the energy specifications with the (safe)
approximated information inferred by the static resource analysis.
Such component is based on our previous work on general resource usage
verification presented
in~\cite{resource-verif-iclp2010-short,resource-verif-2012-short}, where we
extended the criteria of correctness as the conformance of a program
to a specification expressing non-functional global properties, such
as upper and lower bounds on execution time, memory, energy, or user
defined resources, given as functions on input data sizes. We also
defined an abstract semantics for resource usage properties and
operations to compare the (approximated) intended semantics of a
program (i.e., the specification) with approximated semantics inferred
by static analysis. These operations include the comparison of
arithmetic functions (e.g., polynomial, exponential, or logarithmic
functions) that may come from the specifications or from the analysis
results.
As a possible result of the comparison
in the output of the tool,
either:

\begin{enumerate}
 
\item The original (specification) assertion (i.e., with status
  \texttt{check}) is included with status \texttt{checked}
  (resp. \texttt{false}), meaning that the assertion is correct
  (resp. incorrect) for all input data meeting the precondition of the
  assertion,

\item the assertion is ``split'' into two or three assertions with
  different status (\texttt{checked}, \texttt{false}, or
  \texttt{check}) whose preconditions include a conjunct expressing
  that the size of the input data belongs to the interval(s) for which
  the assertion is correct (status \texttt{checked}), incorrect
  (status \texttt{false}), or the tool is not able to determine
  whether the assertion is correct or incorrect (status
  \texttt{check}), or

\item in the worst case, the assertion is included with status
  \texttt{check}, meaning that the tool is not able to prove nor to 
  disprove (any part of) it.

\end{enumerate}

If all assertions are \texttt{checked} then the program is \emph{verified}.
Otherwise, for assertions (or parts of them) that get \texttt{false}
status, a \emph{compile-time error} is reported.  Even if a program
contains no assertions, it can be checked against the assertions
contained in the libraries used by the program, potentially catching
bugs at compile time. Finally, and most importantly, for assertion (or
parts of them) left with status \texttt{check}, the tool can
optionally produce a \emph{verification warning} (also referred to as
an ``alarm''). In addition, optional run-time checks can also be
generated.

\section{The Assertion Language}
\label{sec:assertion-language}

Two aspects of the assertion language are described here: the front-end
language in which assertions are written and included in the XC
programs to be verified, and the internal language in which such
assertions are translated into and passed, together with the \hcir
program representation, to the core analysis and verification tools,
the \ial.

\subsection{The \ciao~Assertion Language}
\label{sec:ciao-assertion-language}
We describe here the subset of the \ciao~assertion language which
allows expressing global ``computational'' properties and, in
particular, resource usage.  
We refer the reader
to~\cite{assert-lang-disciplbook-short,ciaopp-sas03-journal-scp-short,hermenegildo11:ciao-design-tplp-short}
and their references for a full description of this 
assertion language.

For brevity, we only introduce here the class of \texttt{pred}
\textbf{assertions}, which describes a particular predicate and, in
general, follows the schema:\\
\centerline{
\colorbox{lightgrey}{\kbd{:- pred} \nt{Pred} [\kbd{:}
  \nt{Precond\/}]
  [\kbd{=>} \nt{Postcond\/}] [\kbd{+} \nt{Comp-Props\/}]\kbd{.}}
}\\
\noindent
where \nt{Pred} is a predicate symbol applied to distinct free
variables while \nt{Precond} and \nt{Postcond} are logic formulae about
execution states. An execution state is defined by 
variable/value bindings in a given execution step. The assertion indicates that
in any call to \nt{Pred}, if \nt{Precond} holds in the
calling state and the computation of the call succeeds, then
\nt{Postcond} also holds in the success state. Finally, the
\nt{Comp-Props} field is used to describe properties of the whole
computation for calls to predicate \nt{Pred} that meet
\nt{Precond}. In our application \nt{Comp-Props} are precisely the
resource usage properties.

For example, the following assertion for a typical \texttt{append/3} predicate:
\vspace{-1mm}
\begin{lstlisting}
:- pred append(A,B,C) 
   : (list(A,num),list(B,num),var(C)) 
  => (list(C,num),
      rsize(A,list(ALb,AUb,num(ANl,ANu))),
      rsize(B,list(BLb,BUb,num(BNl,BNu))),
      rsize(C,
       list(ALb+BLb,AUb+BUb,
         num(min(ANl,BNl),max(ANu,BNu)))))
   + resource(steps,ALb+1, AUb+1).
\end{lstlisting}

\vspace{-2mm} \noindent
states that for any call to predicate \texttt{append/3}
with the first and second arguments bound to lists of numbers, and the third one
unbound, if the call succeeds, then the third argument will also be bound
to a list of numbers. It also states that
an upper bound on the number of resolution steps required to execute
any of such calls is $AUb + 1$, a function on the length of list $A$. The \texttt{rsize} terms are the \emph{sized types} derived from the regular types, containing variables that represent explicitly lower and upper bounds on the size of terms and subterms appearing in arguments. See Section~\ref{sec:energy-analysis} for an
overview of the general resource analysis framework and how sized types are used.

The global non-functional property \texttt{resource/3} (appearing in the
``\texttt{+}'' field), is used for expressing resource usages and
follows the schema:

\centerline{\colorbox{lightgrey}{\textbf{resource(\nt{Res\_Name}, \nt{Low\_Arith\_Expr, \nt{Upp\_Arith\_Expr})} }}}

\noindent
where \nt{Res\_Name} is a user-provided identifier
for the resource the assertion refers to, \nt{Low\_Arith\_Expr} and \nt{Upp\_Arith\_Expr} are
arithmetic functions that map input data sizes to resource usage, representing respectively lower and upper bounds on the resource consumption.

Each assertion can be in a particular \emph{status}, marked with the
following prefixes, placed just before the \texttt{pred} keyword:
\texttt{check} (indicating the assertion needs to be checked),
\texttt{checked} (it has been checked and proved correct by the
system), \texttt{false} (it has been checked and proved incorrect by
the system; a compile-time error is reported in this case),
\texttt{trust} (it provides information coming from the programmer and
needs to be trusted), or \texttt{true} (it is the result of static
analysis and thus correct, i.e., safely approximated). The default
status (i.e., if no status appears before \texttt{pred}) is
\texttt{check}. 

\subsection{The XC Assertion Language}
\label{sec:xc-assertion-language}

\newcommand\terminal[1]{`{\tt #1}'}

The assertions within XC files use instead a different syntax that is
closer to standard C notation and friendlier for C
developers. These assertions are transparently translated into Ciao
assertions when XC files are loaded into the tool. The Ciao assertions
output by the analysis are also translated back into XC assertions
and added inline to a copy of the original XC file. 

More concretely, the syntax of the XC assertions accepted by our tool is
given by the following grammar, where the non-terminal
\synt{identifier} stands for a standard C identifier, \\ \synt{integer}
stands for a standard C integer, and the non-terminal 
\synt{ground-expr} for a ground expression, i.e., an expression of
type \synt{expr} that does not contain any C identifiers that appear 
in the assertion scope (the non-terminal \synt{scope}).
\begin{grammar}

  <assertion> ::=  
\terminal{\#pragma} <status> <scope> \terminal{:}  <body>

  <status> ::= 
   \terminal{check}   |
   \terminal{trust}   |
   \terminal{true}    | 
   \terminal{checked} |
   \terminal{false}

  <scope> ::= <identifier> \terminal{(}  \terminal{)} 
\alt 
<identifier> \terminal{(} <arguments> \terminal{)}

  <arguments> ::= <identifier> |
  <arguments> \terminal{,}  <identifier>

  <body> ::= <precond>  \terminal{==>} <cost-bounds> | <cost-bounds>

  <precond> ::=  <upper-cond> |   <lower-cond> 
\alt <lower-cond> \terminal{\&\&} <upper-cond>

  <lower-cond> ::= <ground-expr> \terminal{\verb|<=|} <identifier>
  
  <upper-cond> ::=  <identifier> \terminal{\verb|<=|} <ground-expr>

  <cost-bounds> ::= 
   <lower-bound> |  <upper-bound>
  \alt<lower-bound> \terminal{\&\&} <upper-bound>

  <lower-bound> := <expr> \terminal{\verb|<=|} \terminal{energy}

  <upper-bound> := \terminal{energy} \terminal{\verb|<=|}  <expr>
 
  <expr> :=  <expr> \terminal{+} <mult-expr>
  \alt  <expr> \terminal{-} <mult-expr>
  
  <mult-expr> :=  <mult-expr> \terminal{*} <unary-expr>
  \alt  <mult-expr> \terminal{/} <unary-expr>
  
  <unary-expr> := 
 <identifier> 
\alt  <integer>
  \alt \terminal{sum} \terminal{(} <identifier> \terminal{,} <expr> \terminal{,} <expr> \terminal{,} <expr> \terminal{)}
  \alt \terminal{prod} \terminal{(} <identifier> \terminal{,} <expr> \terminal{,} <expr>\terminal{,} <expr> \terminal{)}
  \alt \terminal{power} \terminal{(} <expr> \terminal{,} <expr> \terminal{)} 
  \alt \terminal{log} \terminal{(} <expr> \terminal{,} <expr> \terminal{)} 
  \alt \terminal{(} <expr> \terminal{)}  
  \alt \terminal{+} <unary-expr>
  \alt \terminal{-} <unary-expr>
  \alt \terminal{min} \terminal{(} <identifier> \terminal{)}
  \alt \terminal{max} \terminal{(} <identifier> \terminal{)}
\end{grammar}

XC assertions are  directives starting with the token
{\tt \#pragma} followed by the assertion {\em status}, the assertion
{\em scope}, and the assertion {\em body}.  The assertion {\em
  status} 
can take
several values, 
including \texttt{check}, \texttt{checked}, \texttt{false},
\texttt{trust} or \texttt{true}, with the same meaning as in the Ciao
assertions. Again, the default status is \texttt{check}.

The assertion scope identifies the function the assertion is referring
to, and provides the local names for the arguments of the function to
be used in the body of the assertion.  %
For instance, the scope {\tt biquadCascade(state, xn, N)} refers to the
function {\tt biquadCascade} and binds the arguments within the body
of the assertion to the respective identifiers {\tt state}, {\tt xn},
{\tt N}.  %
While the arguments do not need to be named in a consistent way
w.r.t.\ the function definition, it is highly recommended for the sake
of clarity.  %
The \emph{body} of the assertion expresses bounds on the energy
consumed by the function and optionally contains preconditions (the
left hand side of the {\tt ==>} arrow) that constrain the argument
sizes.

Within the body, expressions of type \synt{expr} are built from
standard integer arithmetic functions (i.e., {\tt +}, {\tt -}, {\tt *},
{\tt /}) plus the following extra functions:
\begin{itemize}
\setlength\itemsep{-4px}
\item {\tt power(base, exp)} is the exponentiation of {\tt base} by {\tt exp};
\item {\tt log(base, expr)} is the logarithm of  {\tt expr} in base {\tt base};
\item {\tt sum(id, lower, upper, expr)} is the summation of the sequence of the values 
   of {\tt expr} for {\tt id} ranging from {\tt lower} to {\tt upper}; 
\item {\tt prod(id, lower, upper, expr)} is the product of the sequence of the values 
   of {\tt expr} for {\tt id} ranging from {\tt lower} to {\tt upper};
 \item {\tt min(arr)} is the minimal value of the array {\tt arr};
\item {\tt max(arr)} is the maximal value of the array {\tt arr}.
\end{itemize}
Note that the argument of {\tt min} and {\tt max} must be an identifier
appearing in the assertion scope that corresponds to an array of integers
(of arbitrary dimension).
  \bigskip

\section{\isa/\llvmir to \hcir Transformation}
\label{sec:llvm-ciao-translation}

In this section we describe briefly the \hcir representation 
and the transformations into it that we developed
in order to achieve the verification tool presented in
Section~\ref{sec:overview} and depicted in
Figure~\ref{fig:analysis-verif-hcir}. The transformation of \isa code
into \hcir was described in~\cite{isa-energy-lopstr13-final}.  
We provide herein an overview of
the \llvmir to \hcir transformation.

The \hcir representation consists of a sequence of \emph{blocks} where
each block is represented as a \emph{Horn clause}:

\centerline{\mbox{$<block\_id>( <params> ) \imp \ S_1, \ \ldots \ ,S_n.$}}

\noindent
Each block has an entry point, that we call the \emph{head} of the
block (to the left of the $\imp$ symbol),
with a number of parameters $<params>$, and a sequence of steps (the
\emph{body}, to the right of the $\imp$ symbol). Each of these $S_i$
steps (or \emph{literals}) is either (the representation of) an
\llvmir \emph{instruction}, or a \emph{call} to another (or the same)
block. The analyzer deals with the \hcir always in the same way,
independent of its origin. 

\llvmir programs are expressed using typed assembly-like
instructions. Each function is in SSA form, represented as a sequence
of basic blocks. Each basic block is a sequence of \llvmir
instructions that are guaranteed to be executed in the same order.
Each block ends in either a branching or a return instruction. In
order to represent each of the basic blocks of the \llvmir in the
\hcir, we follow a similar approach as in 
the ISA-\level transformation~\cite{isa-energy-lopstr13-final}.
However, the \llvmir includes an additional type transformation as
well as better memory modelling. It is explained in detail in Appendix
5 of~\cite{entra-d3.2}.
The main aspects of this process, are the following:

\begin{enumerate}

\item Infer input/output parameters to each block. 

\item Transform \llvmir types into \hcir types.

\item Represent each \llvmir block as an \hcir block and each
  instruction in the \llvmir block as a literal ($S_i$).
  
\item Resolve branching to multiple blocks by creating clauses with
  the same signature (i.e., the same name and arguments in the head),
where each clause denotes one of the blocks the branch may jump to.

\end{enumerate}

The translator component is also in charge of translating the XC
assertions to Ciao assertions and back. 
Assuming the Ciao type of the input and output of the function 
is known,
the translation of assertions from Ciao to XC (and back) is relatively
straightforward.  The \nt{Pred} field of the Ciao assertion is
obtained from the scope of the XC assertion to which an extra argument 
is added representing the output of the function.  The
\nt{Precond} fields are produced directly from the type of the input
arguments: to each input variable, its regular type and its regular
type size are added to the precondition, while the added output argument
is declared as a free variable.  
Finally the \nt{Comp-Props} field is set to the usage of
the resource \texttt{energy}, i.e., a literal of the form
\texttt{resource(energy, Lower, Upper)} where \texttt{Lower} and
\texttt{Upper} are the lower and upper bounds from the energy
consumption specification.

\section{Energy Consumption Analysis}
\label{sec:energy-analysis}

As already mentioned in Section~\ref{sec:overview}, we use an existing
static analysis to infer the energy consumption of XC
programs~\cite{isa-energy-lopstr13-final}. It is a specialization of
the generic resource analysis presented
in~\cite{plai-resources-iclp14} that uses the instruction-level models
described in~\cite{Kerrison13}.  Such generic resource analysis is
fully based on \emph{abstract interpretation}~\cite{Cousot77},
defining the resource analysis itself as an \emph{abstract domain}
that
is integrated into the PLAI abstract interpretation
framework~\cite{ai-jlp,inc-fixp-sas} of \ciaopp, obtaining features
such as \emph{multivariance}, efficient fixpoints, and assertion-based
verification and user interaction. 

In the rest of this section we give an overview of the general
resource analysis, using the following \texttt{append/3} predicate as
a running example:

\begin{small}
\begin{lstlisting}
append([],    S, S).
append([E|R], S, [E|T]) :- append(R,S,T).
\end{lstlisting}
\end{small}

The first step consists of obtaining the regular type of the arguments
for each predicate. To this end, we use one of the type analyses present in the \ciaopp{} system~\cite{eterms-sas02}. In our example, the system infers that for any call to the predicate \texttt{append(X, Y, Z)} with \texttt{X} and \texttt{Y} bound to lists of numbers and
\texttt{Z} a free variable, if the call succeeds, then \texttt{Z} also gets bound to a list of numbers. The regular type for representing ``list of numbers'' is defined as follows:

\begin{lstlisting}
listnum := [] | [num | listnum].
\end{lstlisting}

From this type definition, sized type schemas are derived, which
incorporate variables representing explicitly lower and upper bounds
on the size of terms and subterms.  For example, in the following
sized type schema (named $listnum\text{-}s$):
$$listnum\text{-}s \to listnum^{(\alpha,\beta)}(num^{(\gamma, \delta)})$$

\noindent
$\alpha$ and $\beta$ represent lower and upper bounds on the length of
the list, respectively, while $\gamma$ and $\delta$ represent lower
and upper bounds of the numbers in the list, respectively.

In a subsequent phase, these sized type schemas are put into relation,
producing a system of recurrence equations where output argument sizes
are expressed as functions of input argument sizes.

The resource analysis is in fact an extension of the sized type
analysis that adds recurrence equations for each resource. As the HC
IR representation is a logic program, it is necessary to consider that
a predicate can fail or have more than one solution, so we need an
auxiliary \emph{cardinality analysis} to get more precise results.

We develop the \texttt{append} example for the simple case of the
resource being the number of resolution steps
performed 
by a call to \texttt{append/3} and we will only focus on upper bounds,
$r_U$.
For the first clause, we know that only one resolution step is needed,
so:

\vspace{-2mm}
\begin{small}
$$r_U\left(
  ln^{(0, 0)}(n^{(\gamma_X, \delta_X)}),
  ln^{(\alpha_Y, \beta_Y)}(n^{(\gamma_Y,
    \delta_Y)}) \right) \leq 1 $$
\end{small}
\vspace{-2mm}

\noindent The second clause performs one
resolution step plus all the resolution steps performed by all possible
backtrackings over the call in the body of the clause. This number
can be bounded as a function of the number of solutions. After setting up and solving these equations we infer that an upper bound on the number of resolution steps is the (upper bound on) the length of the input list
\texttt{X} plus one. This is expressed as:

\vspace{-2mm}
\begin{small}
$$r_U\left(
    ln^{(\alpha_X, \beta_X)}(n^{(\gamma_X, \delta_X)}),
    ln^{(\alpha_Y, \beta_Y)}(n^{(\gamma_Y, \delta_Y)}) \right)
  \leq \beta_X + 1$$
\end{small}
We refer the reader to~\cite{plai-resources-iclp14} for a full
description of this analysis and tool.

\section{The General Resource Usage Verification Framework}
\label{sec:resource-verif-framework}

  In this section we describe the general framework for (static)
  resource usage
  \emph{verification}~\cite{resource-verif-iclp2010-short,resource-verif-2012} 
  that
  we have specialized in this paper for verifying energy consumption
  specifications of XC programs.

  The framework, that we introduced in~\cite{resource-verif-iclp2010-short},
  extends the criteria of correctness as the conformance of a program
  to a specification expressing non-functional global properties, such
  as upper and lower bounds on execution time, memory, energy, or user
  defined resources, given as functions on input data sizes.

Both program verification and debugging compare the {\em actual
  semantics} $\sem{P}$ of a program $P$ with an {\em intended
  semantics} for the same program, which we will denote by $\Inten$.
This intended semantics embodies the user's requirements, i.e., it is
an expression of the user's expectations.  In the framework, both
semantics are given in the form of (\emph{safe}) approximations.  The
abstract (safe) approximation $\p_\alpha$ of the concrete semantics
$\p$ of the program is actually computed by (abstract
interpretation-based) \emph{static analyses}, and compared directly to
the (also approximate) specification, which is safely assumed to be
also given as an abstract value $\Inten_\alpha$. Such approximated
specification is expressed by \emph{assertions} in the program.
Program verification is then performed by comparing $\Inten_\alpha$
and $\p_\alpha$.

In this paper, we assume that the program $P$ is in \hcir form (i.e.,
a logic program), which is the result of the transformation of the \isa
or \llvmir code corresponding to an XC program. As already said, such
transformation preserves the resource consumption semantics, in the
sense that the resource usage information inferred by the static
analysis (and hence the result of the verification process) is
applicable to the original XC program.

\paragraph {\bf Resource usage semantics}
Given a program $p$, let ${\cal C}_p$ be the set of all calls to $p$.
The concrete resource usage semantics of a program $p$, for a
particular resource of interest, $\sem{P}$, is a set of pairs $(p(\bar
t), r)$ such that $\bar t$ is a tuple of 
data (either simple data such as numbers, or compound data structures),
$p(\bar t) \in {\cal C}_p$ is a call to 
procedure\footnote{Also called \emph{predicate} in the \hcir.}
$p$ with actual
parameters $\bar t$, and $r$ is a number expressing the amount of
resource usage of the computation of the call $p(\bar t)$.
The concrete resource usage semantics can be defined as a function
$\sem{P}: {\cal C}_p \mapsto {\cal R}$ where ${\cal R}$ is the set of
real numbers (note that depending on the type of resource we can take
other set of numbers, e.g., the set of natural numbers).

The abstract resource usage semantics is a set of 4-tuples:
 $$(p(\bar v):c(\bar v), \Phi, input_p, size_p)$$
where $p(\bar v):c(\bar v)$ is an
 abstraction of a set of calls. $\bar v$ is a tuple of variables and
 $c(\bar v)$ is an abstraction representing a set of tuples of 
 data 
 which are instances of $\bar v$. $c(\bar v)$ is an element of some
 abstract domain expressing instantiation states.
 $\Phi$ is an abstraction of the
 resource usage of the calls represented by $p(\bar v):c(\bar v)$. We
 refer to it as a {\em resource usage interval function} for $p$,
 defined as follows:

\begin{itemize}
\item A {\em resource usage bound function} for $p$ is a monotonic
  arithmetic function, $\Psi: S \mapsto {\cal R}_{\infty}$, for a given
  subset $S \subseteq {\cal R}^{k}$, where ${\cal R}$ is the set of real numbers, $k$ is the number of
  input arguments to 
procedure 
$p$ and ${\cal R}_{\infty}$ is the set
  of real numbers augmented with the special symbols $\infty$ and
  $-\infty$. We use such functions to express lower and upper bounds
  on the resource usage of 
procedure
$p$ depending on input data
  sizes.

\item A {\em resource usage interval function} for $p$ is an
  arithmetic function, $\Phi: S \mapsto {\cal RI}$, where
  $S$ is defined as before and ${\cal RI}$ is the set of intervals
  of real numbers, such that $\Phi(\bar n) = [\Phi^l(\bar n),
  \Phi^u(\bar n)]$ for all $\bar n \in S$, where
  $\Phi^l(\bar n)$ and $\Phi^u(\bar n)$ are {\em resource usage bound
    functions} that denote the lower and upper endpoints of the
  interval $\Phi(\bar n)$ respectively for the tuple of input data
  sizes $\bar n$.  Although $\bar n$ is typically a tuple of natural
  numbers, we do not want to restrict our framework.  We require that
  $\Phi$ be well defined so that $\forall \bar n \ (\Phi^l(\bar n)
  \leq \Phi^u(\bar n))$.
\end{itemize}

\noindent
$input_p$ is a function that takes a tuple of 
data $\bar t$ and returns a tuple with the input arguments to
$p$. This function can be inferred by using the existing mode analysis or
be given by the user by means of assertions.  $size_p(\bar t)$ is
a function that takes a tuple of terms $\bar t$ and returns a tuple
with the sizes of those data under the size metric described in
Section~\ref{sec:energy-analysis}.

In order to make the presentation simpler, we will omit the $input_p$
and $size_p$ functions in abstract tuples, with the understanding that
they are present in all such tuples.

\paragraph{\bf Intended meaning}

The intended approximated meaning $\Inten_\alpha$ of a program is an
abstract semantic object with the same kind of tuples: $(p(\bar
v):c(\bar v), \Phi,input_p, size_p)$, which
is represented by using \ciao assertions (which are part of the \hcir)
of the form: \\ 
\centerline{
\colorbox{lightgrey}{\kbd{:- check} \nt{Pred} [\kbd{:}
    \nt{Precond}\ ] \kbd{+} \nt{ResUsage}\kbd{.}}\\
}\\
\noindent
where $p(\bar v):c(\bar v)$ is defined by \nt{Pred} and \nt{Precond},
and $\Phi$ is defined by \nt{ResUsage}.  The information about
$input_p$ and $size_p$ is implicit in \nt{Precond} and \nt{ResUsage}.
The concretization of $\Inten_\alpha$, $\gamma(\Inten_\alpha)$, is the
set of all pairs $(p(\bar t), r)$ such that $\bar t$ is a tuple of
terms and $p(\bar t)$ is an instance of \nt{Pred} that meets
precondition \nt{Precond}, and $r$ is a number that meets the
condition expressed by \nt{ResUsage} (i.e., $r$ lies in the interval
defined by \nt{ResUsage}) for some assertion.

\begin{Example}
\label{ex:nrevcomp}
Consider the following \hcir program that computes the factorial of
an integer.

\begin{lstlisting}[language=ciao]
fact(N,Fact) :- N=<0, Fact=1.
fact(N,Fact) :- N>0, N1 is N-1,
        fact(N1,Fact1), Fact is N*Fact1.
\end{lstlisting}
One could use the assertion:
\begin{lstlisting}[language=ciao]
:- check pred fact(N,F)  
   : (num(N), var(F)) 
  => (num(N), num(F), 
      rsize(N, num(Nmin, Nmax)), 
      rsize(F, num(Fmin, Fmax)))
   + resource(steps, Nmin+1, Nmax+1).
 \end{lstlisting}

\noindent
to express that for any call to \texttt{fact(N,F)} with the first
argument bound to a number and the second one a free variable, the
number of resolution (execution) steps performed by the computation is
always between $\texttt{Nmin}+1$ and $\texttt{Nmax}+1$, where
\texttt{Nmin} and \texttt{Nmax} respectively stand for a lower and an
upper bound of \texttt{N}. In this concrete example, the lower and
upper bounds are the same, i.e., the number of resolution steps is
exactly $\texttt{N}+1$, but note that they could be different.
\end{Example}

\begin{Example}
The assertion in Example~\ref{ex:nrevcomp}
captures the following concrete semantic tuples:
\begin{center}
\texttt{( fact(0, Y), 1 )}  \qquad  \texttt{( fact(8, Y), 9 )}      
\end{center}
but it does not capture the following ones:
\begin{center}
\texttt{( fact(N, Y), 1 )}  \qquad  \texttt{( fact(1, Y), 35 )}      
\end{center}
the left one in the first line above because it is outside the scope of
the assertion (i.e., \text{N} being a variable, it does not meet the
precondition \nt{Precond}), and the right one because it violates the
assertion (i.e., it meets the precondition \nt{Precond}, but does not
meet the condition expressed by \nt{ResUsage}).
\end{Example}

\paragraph{\bf Partial correctness: comparing to the abstract semantics}

Given a program $p$ and an intended resource usage semantics $\Inten$,
where 
$\Inten: {\cal C}_p  \mapsto {\cal R}$, 
we say that $p$
is partially correct w.r.t.\ $\Inten$ if for all $p(\bar t) \in {\cal
  C}_p$ we have that $(p(\bar t), r) \in \Inten$, where $r$ is
precisely the amount of resource usage of the computation of the call
$p(\bar t)$.  We say that $p$ is partially correct with respect to a
tuple of the form $(p(\bar v):c_I(\bar v), \Phi_I)$ if for all $p(\bar
t) \in {\cal C}_p$ such that $r$ is the amount of resource usage of
the computation of the call $p(\bar t)$, it holds that: if $p(\bar t)
\in \gamma(p(\bar v):c_I(\bar v))$ then $r \in \Phi_I(\bar s)$, where
$\bar s = size_{p}(input_{p}(\bar t))$.  Finally, we say that $p$ is
partially correct with respect to $\Inten_{\alpha}$ if:
\begin{itemize}
\item For all $p(\bar t) \in {\cal C}_p$, there is a tuple $(p(\bar
  v):c_I(\bar v), \Phi_I)$ in $\Inten_{\alpha}$ such that $p(\bar t)
  \in \gamma(p(\bar v):c_I(\bar v))$, and

\item $p$ is partially correct with respect to every tuple in $\Inten_{\alpha}$.
\end{itemize}

Let $(p(\bar v):c(\bar v), \Phi)$ and $(p(\bar v):c_I(\bar v),
\Phi_I)$ be tuples expressing an abstract semantics $\p_\alpha$
inferred by analysis and an intended abstract semantics
$\Inten_\alpha$, respectively, such that $c_I(\bar v) \sqsubseteq
c(\bar v)$,\footnote{Note that the condition $c_I(\bar v) \sqsubseteq
  c(\bar v)$ can be checked using the \ciaopp~capabilities for
  comparing program state properties such as types.}
and for all $\bar n \in S$ ($S \subseteq {\cal R}^{k}$), $\Phi(\bar n) = [\Phi^l(\bar
n), \Phi^u(\bar n)]$ and $\Phi_I(\bar n) = [\Phi_I^l(\bar n),
\Phi_I^u(\bar n)]$.  We have that:

\begin{itemize} 
\item[(1)] If for all $\bar n \in S$, $\Phi_{I}^{l}(\bar n) \leq
  \Phi^{l}(\bar n)$ and $\Phi^{u}(\bar n) \leq \Phi_{I}^{u}(\bar n)$,
  then $p$ is partially correct with respect to $(p(\bar v):c_I(\bar
  v), \Phi_I)$.

\item[(2)] If for all $\bar n \in S$ $\Phi^{u}(\bar n) <
  \Phi_{I}^{l}(\bar n)$ or $\Phi_{I}^{u}(\bar n) < \Phi^{l}(\bar n)$,
  then $p$ is incorrect with respect to $(p(\bar v):c_I(\bar v),
  \Phi_I)$.

\label{conditions}
\end{itemize}

Checking the two conditions above requires the comparison of resource
usage bound functions.

\paragraph{\bf Resource Usage Bound Function Comparison}

Since the resource analysis we use is able to infer different types of
functions (e.g., polynomial, exponential, and logarithmic), it is also
desirable to be able to compare all of these functions.

For simplicity of exposition, consider first the case where resource
usage bound functions depend on one argument.  Given two resource
usage bound functions (one of them inferred by the static analysis and
the other one given in an assertion/specification present in the
program), $\Psi_{1}(n)$ and $\Psi_{2}(n)$, $n \in {\cal R}$ the
objective of the comparison operation is to determine intervals for
$n$ in which $\Psi_{1}(n) > \Psi_{2}(n)$, $\Psi_{1}(n) = \Psi_{2}(n)$,
or $\Psi_{1}(n) < \Psi_{2}(n)$. For this, we define $f(n) =
\Psi_{1}(n)-\Psi_{2}(n)$ and find the roots of the equation
$f(n)=0$. Assume that the equation has $m$ roots, $n_1, \ldots, n_m$.
These roots are intersection points of $\Psi_{1}(n)$ and
$\Psi_{2}(n)$.  We consider the intervals $S_1 = [0, n_1)$, $S_2 =
  (n_1, n_2)$, $S_m = \ldots$ $(n_{m-1}, n_m)$, $S_{m+1} = (n_m,
  \infty)$. For each interval $S_i$, $1 \leq i \leq m$, we select a
  value $v_i$ in the interval. If $f(v_i)>0$ (respectively
  $f(v_i)<0$), then $\Psi_{1}(n) > \Psi_{2}(n)$ (respectively
  $\Psi_{1}(n) < \Psi_{2}(n)$) for all $n \in S_i$.

There exist powerful algorithms for obtaining roots of polynomial
functions. In our implementation we have used the GNU Scientific
Library~\cite{gsl-short}, which offers a specific polynomial function
library that uses analytical methods for finding roots of polynomials
up to order four, and uses numerical methods for higher order
polynomials.

We approximate exponential and logarithmic resource usage functions
using Taylor series.  In particular, for exponential functions we use
the following formulae:
\[e^x\approx \Sigma_{n=0}^\infty \frac{x^n}{n!}=1+x+\frac{x^2}{2!}+
\frac{x^3}{3!}+\ldots \qquad for\ all\ x\]
\[a^x=e^{x\ ln\ a}\approx 1+x\ ln\ a+\frac{(x\ ln\ a)^2}{2!}+ 
\frac{(x\ ln\ a)^3}{3!}+\ldots\]

\noindent
In our implementation these series are limited up to order 8. This
decision has been taken based on experiments we have carried out that
show that higher orders do not bring a significant difference in
practice.  Also, in our implementation, the computation of the
factorials is done separately and the results are kept in a table in
order to reuse them.

Dealing with logarithmic functions is more complex, as Taylor series
for such functions can only be defined for the interval $(-1, 1)$.

For resource usage functions depending on more than one variable, the
comparison is performed using constraint solving techniques.

\paragraph{\bf Safety of the Approximations}

When the roots obtained for function comparison are approximations of
the actual roots, we must guarantee that their values are safe, i.e.,
that they can be used for verification purposes, in particular, for
safely checking the conditions presented above.  In other words, we
should guarantee that the error falls on the safe side when comparing
the corresponding resource usage bound functions. For this purpose we
developed an algorithm for detecting whether the approximated root
falls on the safe side or not, and in the case it does not fall on the
safe side, performing an iterative process to increment (or decrement)
it by a small value until the approximated root falls on the safe
side.

\section{Using the Tool: Example}

\begin{figure*}[t]
\centerline{\includegraphics[width=\textwidth]{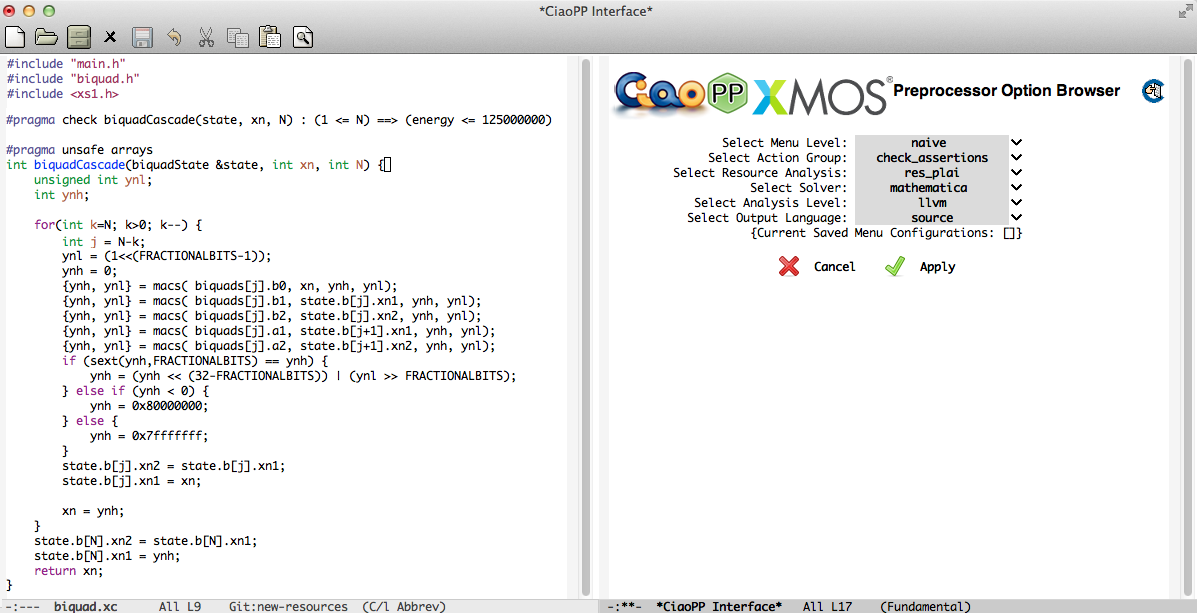}}
\caption{Graphical User Interface of the prototype with the XC biquad program.}
\label{fig:biquadmenu}
\end{figure*}

As an illustrative example of a scenario where the embedded software
developer has to decide values for program parameters that meet an
energy budget, we consider the development of an equaliser (XC)
program using a biquad filter. In Figure~\ref{fig:biquadmenu} we can
see what the graphical user interface of our prototype looks like,
with the code of this biquad example ready to be verified. The purpose
of an equaliser is to take a signal, and to attenuate / amplify
different frequency bands. For example, in the case of an audio
signal, this can be used to correct for a speaker or microphone
frequency response. The energy consumed by such
a program directly depends on several parameters, such as the sample
rate of the signal, and the number of banks (typically between 3 and
30 for an audio equaliser). A higher number of banks enables the
designer to create more precise frequency response curves.

Assume that the developer has to decide how many banks to use in order
to meet an energy budget while maximizing the precision of frequency
response curves at the same time.
In this example, the developer writes an XC program where the number of banks
is a variable, say $\texttt{N}$. Assume also that the energy constraint to be
met is that an application of the biquad program should consume less
than 125 millijoules (i.e., 125000000 nanojoules). 
This constraint is expressed by the following check assertion (specification):
\begin{verbatim}
#pragma check biquadCascade(state,xn,N) : 
   (1 <= N) ==> (energy <= 125000000)
\end{verbatim}

\noindent
where the precondition $\texttt{1 <= N}$ in the assertion (left hand
side of $\texttt{==>}$) expresses that the number of banks should be
at least 1.

Then, the developer makes use of the tool, by selecting the following
menu options, as shown in the right hand side of
Figure~\ref{fig:biquadmenu}: \texttt{check\_assertions}, for
\texttt{Action Group}, \texttt{res\_plai}, for \texttt{Resource
  Analysis}, \texttt{mathematica}, for \texttt{Solver}, \texttt{llvm},
for \texttt{Analysis Level} (which will tell the analysis to take the
\llvmir option by compiling the source code into \llvmir and transform
into \hcir for analysis) and finally \texttt{source}, for
\texttt{Output Language} (the language in which the analysis /
verification results are shown). After clicking on the \texttt{Apply}
button below the menu options, the analysis is performed,
which infers a lower and an upper bound function for the consumption
of the program.  Concretely those bounds are represented by the
following assertion, which is included in the output of the tool:
\begin{verbatim}
#pragma true biquadCascade(state,xn,N) : 
   (16502087*N + 5445103 <= energy && 
    energy <= 16502087*N + 5445103)
\end{verbatim}

In this particular case, both bounds are identical. In other words,
the energy consumed by the program is exactly characterized by the
following function, depending on $\texttt{N}$ only:
\[
E_\text{biquad}(\texttt{N}) = 16502087 \times  \texttt{N} + 5445103 \text{ nJ}
\]

 Then, the verification of the specification (check assertion) is
 performed by comparing the energy bound functions above with the
 upper bound expressed in the specification, i.e., 125000000, a
 constant value in this case. As a result, the two following
 assertions are produced (and included in the output file of the
 tool):

\begin{verbatim}
#pragma checked biquadCascade(state,xn,N) : 
   (1 <= N && N <= 7) 
       ==> (energy <= 125000000)
#pragma false biquadCascade(state,xn,N) : 
   (8 <= N) 
       ==> (energy <= 125000000)
\end{verbatim}

The first one expresses that the original assertion holds subject to a
precondition on the parameter $\texttt{N}$, i.e., in order to meet the
energy budget of 125 millijoules,
the number of banks $\texttt{N}$ should be a natural number in the
interval $[1,~7]$ (precondition $\texttt{1 <= N \&\& N <= 7}$). The
second one expresses that the original specification is not met
(status \texttt{false}) if the number of banks is greater or equal to
$8$.

 Since the goal is to maximize the precision of frequency response
 curves and to meet the energy budget at the same time, the number of
 banks should be set to 7. The developer could also be interested in
 meeting an energy budget but this time ensuring a lower bound on the
 precision of frequency response curves. For example by ensuring that
 $\texttt{N} \geq 3$, the acceptable values for $\texttt{N}$ would be
 in the range $[3,~7]$.

 In the more general case where the energy function inferred by the
 tool depends on more than one parameter, the determination of the
 values for such parameters is reduced to a constraint solving
 problem. The advantage of this approach is that the parameters can be
 determined analytically at the program development phase, without the
 need of determining them experimentally by measuring the energy of
 expensive program runs with different input parameters.

\ \\

\section{Related Work}

As mentioned before, this work adds verification capabilities to our
previous work on energy consumption analysis for
XC/XS1-L~\cite{isa-energy-lopstr13-final}, which builds on of our
general framework for resource usage
analysis~\cite{resource-iclp07,NMHLFM08,plai-resources-iclp14,ciaopp-sas03-journal-scp,decomp-oo-prolog-lopstr07}
and its support for resource
verification~\cite{resource-verif-iclp2010-short,resource-verif-2012}, 
and the energy models of~\cite{Kerrison13}. 

Regarding the support for verification of properties expressed as
functions, the closest related work we are aware of
presents a method for comparison of cost functions inferred by the
COSTA system for Java bytecode~\cite{AlbertAGHP09}. The method proves
whether a cost function is smaller than another one \emph{for all the
  values} of a given initial set of input data sizes.  The result of
this comparison is a Boolean value. However, as mentioned before, in
our approach~\cite{resource-verif-iclp2010-short,resource-verif-2012} the result is in general a set
of subsets (intervals) in which the initial set of input data sizes is
partitioned, so that the result of the comparison is different for
each subset.  Also, \cite{AlbertAGHP09} differs in that comparison is
syntactic, using a method similar to what was already being done in
the~\ciaopp~system: performing a function normalization and then using
some syntactic comparison rules.
Our technique goes beyond
these syntactic comparison rules. Moreover,~\cite{AlbertAGHP09} only
covers (generic) cost function comparisons while we have addressed the
whole process for the case of energy consumption \emph{verification}.
Note also that, although we have presented our work applied to XC
programs, the CiaoPP system can also deal with other high- and
low-level languages, including, e.g., Java
bytecode~\cite{resources-bytecode09-short,decomp-oo-prolog-lopstr07}.

In a more general context, using abstract interpretation in debugging
and/or verification tasks has now become well established. To cite
some early work, abstractions were used in the context of algorithmic
debugging in \cite{Lichtenstein88}.  Abstract interpretation has been
applied by Bourdoncle \cite{Bourd} to debugging of imperative programs
and by Comini et al.\ to the algorithmic debugging of logic programs
\cite{CominiDD95} (making use of partial specifications in
\cite{abs-diag-jlp}), and by P.\ Cousot~\cite{Cousot-VMCAI03} to
verification, among others.  The
\ciaopp~framework~\cite{aadebug97-informal,prog-glob-an,ciaopp-sas03-journal-scp}
was pioneering in many aspects, offering an integrated approach combining
abstraction-based verification, debugging, and run-time checking with
an assertion language.

\section{Conclusions}

We have specialized an existing general framework for resource usage
verification for verifying energy consumption specifications of
embedded programs.  These specifications can include both lower and
upper bounds on energy usage, expressed as intervals within which the
energy usage is supposed to be included, the bounds (end points of the
intervals) being expressed as functions on input data sizes.  Our tool
can deal with different types of energy functions (e.g., polynomial,
exponential or logarithmic functions), in the sense that the analysis
can infer them, and the specifications can involve them. We have shown
through an example, and using the prototype implementation of our
approach within the \ciao/\ciaopp~system and for the XC language and
XS1-L architecture, how our verification system can prove whether such
energy usage specifications are met or not, or infer particular
conditions under which the specifications hold. These conditions are
expressed as intervals of input data sizes such that a given
specification can be proved for some intervals but disproved for
others. The specifications themselves can also include preconditions
expressing intervals for input data sizes.  We have illustrated
through this example how embedded software developers can use this
tool, 
and in particular for determining values for program parameters that
ensure meeting a given energy budget while minimizing the loss in
quality of service.

\section{Acknowledgements}

The research leading to these results has received funding from the
European Union 7th Framework Programme under grant agreement 318337,
ENTRA - Whole-Systems Energy Transparency, Spanish MINECO TIN'12-39391
\emph{StrongSoft} and TIN'08-05624 \emph{DOVES} projects, and Madrid
TIC-\-1465 \emph{PROMETIDOS-CM} project. We also thank all the
participants of the ENTRA project team, and in particular John P.
Gallagher, Henk Muller, Kyriakos Georgiou, Steve Kerrison, and Kerstin
Eder for useful and fruitful discussions. Henk Muller (XMOS Ltd.) also
provided 
benchmarks (e.g., the biquad program) that we used to test our tool.

\bibliographystyle{abbrv}

\end{document}